\documentclass[conference]{IEEEtran}
\IEEEoverridecommandlockouts
\usepackage{cite}
\usepackage{amsmath,amssymb,amsfonts}
\usepackage{algorithmic}
\usepackage{graphicx}
\usepackage{textcomp}
\usepackage{xcolor}
\def\BibTeX{{\rm B\kern-.05em{\sc i\kern-.025em b}\kern-.08em
    T\kern-.1667em\lower.7ex\hbox{E}\kern-.125emX}}

\usepackage{algorithm}
\usepackage{enumitem}
\usepackage{dblfloatfix}
\usepackage{lipsum}
\usepackage{soul}
\usepackage{verbatimbox,caption,float,lipsum}
\newfloat{Code}
\usepackage{algpseudocode}
\usepackage{wrapfig}
\usepackage{gensymb}
\usepackage{textcomp}
\usepackage{algorithm}
\usepackage{listings}
\usepackage{subcaption}
\usepackage{booktabs}
\usepackage{amsmath}
\usepackage{enumitem}
\usepackage{dblfloatfix}
\usepackage{lipsum}
\usepackage{soul}
\usepackage{xcolor}
\usepackage{url}
\usepackage{algorithmic}
\usepackage{textcomp}
\usepackage[table]{xcolor}
\definecolor{asparagus}{rgb}{0.53, 0.66, 0.42}
\definecolor{azure}{rgb}{0.0, 0.5, 1.0}
\definecolor{darkgreen}{rgb}{0.0, 0.5, 0.0}
\definecolor{amaranth}{rgb}{0.9, 0.17, 0.31}
\definecolor{cadetgrey}{rgb}{0.57, 0.64, 0.69}
\definecolor{aureolin}{rgb}{0.99, 0.93, 0.0}
\usepackage{amsfonts}
\usepackage{float}
\usepackage{relsize}
\usepackage{grffile}
\usepackage{algorithm}
\usepackage{algorithmic}
\usepackage{csquotes}
\usepackage[most]{tcolorbox}
\usepackage{listings}
\usepackage{pgfplots}
\usetikzlibrary{arrows.meta,positioning}

\newcommand{\tikzxmark}{%
\tikz[scale=0.23] {
    \draw[line width=0.7,line cap=round] (0,0) to [bend left=6] (1,1);
    \draw[line width=0.7,line cap=round] (0.2,0.95) to [bend right=3] (0.8,0.05);
}}

\newcommand{\cpp}[0]{C\texttt{++}}
\newcommand{\blackcirclednumber}[1]{%
    \tikz[baseline=(X.base)] \node[draw,circle,fill=black,text=white,inner sep=0.8pt] (X) {#1};%
}
\begin{document}

\title{LLM-HPC\texttt{++}: Evaluating LLM-Generated Modern C\texttt{++} and MPI+OpenMP Codes for Scalable Mandelbrot Set Computation}

\author{\IEEEauthorblockN{Patrick Diehl}
\IEEEauthorblockA{\textit{Los Alamos National Laboratory} \\
\textit{Computing and Artificial Intelligence}\\
Los Alamos, NM, USA \\
diehlpk@lanl.gov }% or ORCID}
\and
\IEEEauthorblockN{Noujoud Nader}
\IEEEauthorblockA{\textit{Louisiana State University} \\
\textit{Center for Computation and Technology}\\
Baton Rouge, LA, USA \\
nnader@lsu.edu}% or ORCID}
\and
\IEEEauthorblockN{Deepti Gupta}
\IEEEauthorblockA{\textit{Texas A\&M University-Central Texas} \\
\textit{Computer Information Systems}\\
Killeen, TX, USA\\
d.gupta@tamuct.edu}
% \and
% \and
% \and
}

\maketitle

\begin{abstract}
Parallel programming remains one of the most challenging aspects of High-Performance Computing (HPC), requiring deep knowledge of synchronization, communication, and memory models. While modern C++ standards and frameworks like OpenMP and MPI have simplified parallelism, mastering these paradigms is still complex. Recently, Large Language Models (LLMs) have shown promise in automating code generation, but their effectiveness in producing correct and efficient HPC code is not well understood. In this work, we systematically evaluate leading LLMs including ChatGPT 4 and 5, Claude, and LLaMA on the task of generating C++ implementations of the Mandelbrot set using shared-memory (asynchronous programming, parallel algorithms, coroutines, and OpenMP), and distributed-memory (MPI) paradigms. Each generated program is compiled and executed with GCC 11.5.0 to assess its correctness, robustness, and scalability. Results show that ChatGPT-4 and ChatGPT-5 achieve strong syntactic precision and scalable performance.
\end{abstract}


\section{Introduction}
High-Performance Computing (HPC) relies heavily on efficient parallel programming models to achieve scalability across shared and distributed memory systems \cite{robey2021parallel}. During the past decade, the evolution of modern C++ standards, from asynchronous programming in C++11 \cite{drocco2017parallel}, to parallel algorithms in C++17 \cite{lin2022evaluating}, and coroutines in C++20 \cite{belson2020c}, has significantly improved the ability of developers to express parallelism natively within the language \cite{diehl2024parallel}. Simultaneously, directive-based and message-passing models such as OpenMP and MPI continue to serve as foundational tools for exploiting multi-core and distributed architectures \cite{kang2015performance}. Despite these advances, mastering parallel programming remains complex, requiring a deep understanding of synchronization, memory hierarchy, and communication overheads \cite{gropp1999using}. In the current scenario, Large Language Models (LLMs) have been used to automate code generation and help developers perform high-performance computing tasks \cite{diehl2024evaluating, diehl2025llm, nader2025llm}. Models such as ChatGPT\cite{achiam2023gpt}, Claude \cite{claudeClaude}, and LLaMA \cite{touvron2023LLaMA} have demonstrated strong capabilities in synthesizing functional code from natural language instructions. However, the effectiveness of LLMs in generating correct, efficient, and portable parallel programs, particularly in C++, remains largely unexamined \cite{nader2025llm}. Since parallel programming involves subtle coordination between threads, processes, and memory models, even minor code inaccuracies can lead to build failures, race conditions, or incorrect numerical results \cite{diaz2012survey}. Understanding how LLMs handle these challenges is essential for assessing their utility in HPC development pipelines. 

This study presents a systematic evaluation of multiple state-of-the-art LLMs for generating parallel and distributed C++ codes for computing the Mandelbrot set—a canonical benchmark in scientific visualization and computational workload balancing. The Mandelbrot set is deliberately chosen here as a controlled benchmark \cite{drakopoulos2003overview}. Its embarrassingly parallel nature allows us to isolate and evaluate the LLMs' ability to correctly generate, compile, and debug parallel code across multiple paradigms—C++11/17/20, OpenMP, and MPI+OpenMP—without confounding the assessment with intricate synchronization or communication logic that would obscure the core focus of this study. 

Using a series of carefully designed prompts, we test the LLMs' ability to produce code implementing: \blackcirclednumber{1} Shared-memory parallelism via C++11 asynchronous features, C++17 parallel algorithms, and C++20 coroutines. \blackcirclednumber{2} Directive-based parallelism using OpenMP. \blackcirclednumber{3} Distributed-memory parallelism combining MPI with OpenMP for hybrid execution. Notably, C++20 coroutines represent the most recent standardized mechanism for expressing cooperative concurrency in C++, and their adoption is actively being integrated into modern HPC frameworks such as Kokkos and HPX, making their evaluation in the context of LLM-generated code particularly timely and novel. Each generated program was compiled using GCC 11.5.0, and all build and runtime errors were systematically documented and corrected to evaluate the LLMs' debugging robustness and error patterns. The resulting Portable Bitmap (PBM) visual outputs were validated both visually and functionally to ensure correctness. This process allowed us to assess the reliability, syntactic precision, and semantic consistency of the LLM-generated C++ code across varying levels of parallel abstraction. The main contribution of the paper as follows:
\begin{itemize}
    \item We design structured prompts targeting key paradigms of modern parallelism—C++11/17/20, OpenMP, and MPI+OpenMP—and use them consistently across multiple LLMs.

    \item We provide a detailed analysis of compilation outcomes, runtime errors, and functional correctness of generated codes.

    \item We evaluate LLM competence on C++20 coroutines—the newest standardized parallel abstraction in C++—which are increasingly adopted by cutting-edge HPC frameworks such as Kokkos and HPX.
    
    \item We document and categorize build-time and runtime errors, offering insights into LLM limitations in HPC code synthesis and debugging.
\end{itemize}

Through this comprehensive evaluation, our work bridges the gap between AI-assisted code generation and high-performance C++ programming, providing a first-of-its-kind analysis of how current LLMs perform when tasked with generating complex, parallelizable scientific codes. 
The paper is structured as follows: Section~\ref{sec:related:work} shows the related work. Section~\ref{sec:parallel:cpp} introduces built-in functionality in the C\texttt{++} standard for parallelism. The Mandelbrot set as the example problem to generate an artificial workload is presented in Section~\ref{sec:mandelbrot}. Section~\ref{sec:prompt:design} lists the used prompts. Section \ref{sec:results} presents the results while Section~\ref{sec:code:quality} analyzes the quality of the generated codes. Section~\ref{sec:scaling} presents the scaling results, and finally Section~\ref{sec:conclusion} concludes the paper and discusses future work.

\begin{figure*}[]
    \centering
\begin{tikzpicture}[x=3.5cm,y=1cm]
  % Timeline
  \draw[thick,-{Stealth[length=3mm]}] (-0.3,0) -- (3.4,0);
  \foreach \x in {0,1,2,3} { \draw[thick] (\x,0.12) -- (\x,-0.12); }

  % C++11
  \node[above=2mm] at (0,0) {C++ 11};
  \node[below=2mm,align=left] at (0,0) {%
    \texttt{std::thread}\\
    \texttt{std::async}\\
    Smart pointer\\
    Lambda functions
  };

  % C++14
  \node[above=2mm] at (1,0) {C++ 14};
  \node[below=2mm,align=left] at (1,0) {%
    Generic lambda\\
    shared mutex
  };

  % C++17
  \node[above=2mm] at (2,0) {C++ 17};
  \node[below=2mm,align=left] at (2,0) {%
    Parallel\\
    algorithms
  };

  % C++20
  \node[above=2mm] at (3,0) {C++ 20};
  \node[below=2mm,align=left] at (3,0) {%
    Coroutines\\
    Ranges\\
    Semaphores\\
    Latch\\
    Barrier
  };
\end{tikzpicture}
    \caption{Time line for the features added to the C\texttt{++} standard with respect to parallelism. Adapted from~\cite{10.1007/978-3-031-32316-4_3}.}
    \label{fig:cpp:standard}
\end{figure*}
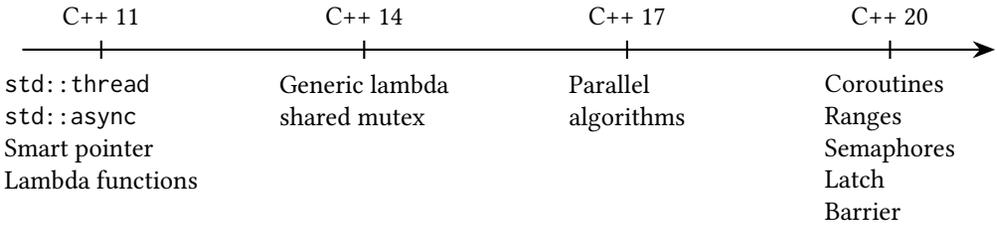

\section{Related work}
\label{sec:related:work}
Several studies have benchmarked LLMs for code generation tasks \cite{diehl2024evaluating,diehl2025llm,chen2021evaluating}, yet their use in analyzing and optimizing HPC workloads remains limited. LLM4HPC \cite{chen2023lm4hpc} was the first framework tailored for HPC, demonstrating promising results in code similarity analysis, parallelism detection, and OpenMP question answering. Building on this, LLM4VV \cite{munley2024llm4vv} fine-tuned GPT-4 and CodeLLaMA (based on LLaMA-2) to generate OpenACC directives effectively. HPC-GPT \cite{ding2023hpc}, developed from LLaMA-13B, has been applied to AI model management, dataset handling, and race detection. Tokompiler \cite{kadosh2023scope}, a domain-specific LLM, outperformed GPT-3-based models for Fortran, C, and C\text{++} code completion. HPC-Coder in both versions further explored code completion for OpenMP, MPI and CUDA constructs \cite{nichols2024hpc, chaturvedi2025hpc}. Lei et al. \cite{lei2023creating} released a dataset for fine-tuning models on OpenMP Fortran and \cpp translation, achieving results surpassing GPT-4. Chen et al. \cite{chen2024landscape} provided a broad overview of LLM–HPC integration challenges beyond code generation. Additionally, Godoy et al. \cite{godoy2023evaluation} and Valero-Lara et al. \cite{valero2023comparing} evaluated LLaMA-2-based HPC kernel generation, while Godoy et al. \cite{godoy2024large} applied GPT-3 for code generation and auto-parallelization in \cpp, Fortran, Python, and Julia. Most recently, Yin et al. \cite{yin2025chathpc} introduced chatHPC, a chatbot supporting HPC question answering and script generation. Schneider et al. \cite{schneider2024mpirigen} introduced the
MPIrigen model approach for generating MPI code.  Chen et al. \cite{chen2024ompgpt} introduced OMPGPT for generating OpenMP
code.

\section{Parallelism within the C\texttt{++} standard}
\label{sec:parallel:cpp}
With the C\texttt{++} 11 standard, \lstinline[language=c++]{std::thread} and \lstinline[language=c++]{std::async} were added as well for parallelism. Lambda functions and smart pointers are utilities. The C\texttt{++} 14 standard added generic lambda functions and a shared mutex. With the C\texttt{++} 17 standard, the parallel algorithms were introduced. With the C\texttt{++} 20 standard, coroutines were added~\cite{10.1007/978-3-031-32316-4_3}. As utilities, the following were added: ranges, semaphores, latch, and barriers. Figure~\ref{fig:cpp:standard} sketches the timeline. All of the built-in C\texttt{++} features for parallelism will be investigated for LLM-based generated codes. 

\subsection{Asynchronous programming}
With the C\texttt{++} 11 standard, the header \lstinline[language=c++]{#include <future>}\footnote{\url{https://en.cppreference.com/w/cpp/thread/future.html}} was introduced for asynchronous programming. With \lstinline[language=c++]{std::async}, a function or lambda function can be executed asynchronously on a dedicated core. Each function returns a promise \lstinline[language=c++]{std::future<T>}, namely a future which contains the return type of \lstinline[language=c++]{T} when the computation finishes. Once the result is ready, it can be obtained using the \lstinline[language=c++]{.get()} function; if the result is not ready yet, the code blocks and waits until the result is ready. For a detailed implementation, we refer to~\cite[Chapter 9]{diehl2024parallel}.

\subsection{Parallel algorithms}
With the C\texttt{++} 17 standard, the algorithms provided by the algorithm library\footnote{\url{https://en.cppreference.com/w/cpp/algorithm.html}} were extended with execution policies. The header \lstinline[language=c++]{#include <execution>}\footnote{\url{https://en.cppreference.com/w/cpp/header/execution.html}} provides the following execution policies: \lstinline[language=c++]{std::execution::par} for parallel execution, \lstinline[language=c++]{std::execution::seq} for serial execution, and \lstinline[language=c++]{std::execution::par_unseq} for parallel execution with potential vectorization optimizations.

\subsection{Coroutines}
With the C\texttt{++} 20 standard the header, \lstinline[language=c++]{#include <coroutine>}\footnote{\url{https://en.cppreference.com/w/cpp/language/coroutines.html}} introduces a stackless, suspendable function which can pause and restart at well-defined states. Only language keywords, like \lstinline[language=c++]{co_await}, \lstinline[language=c++]{co_yield}, and \lstinline[language=c++]{co_return}, are available. However, task types and schedulers need to be implemented. For a detailed implementation, we refer to~\cite[Chapter 11]{diehl2024parallel}.

\section{Example problem: Mandelbrot set}
\label{sec:mandelbrot}
\begin{figure*}[tb]
    \centering
\begin{subfigure}[t]{.35\linewidth}
    \begin{center}
    \includegraphics[width=\linewidth]
    {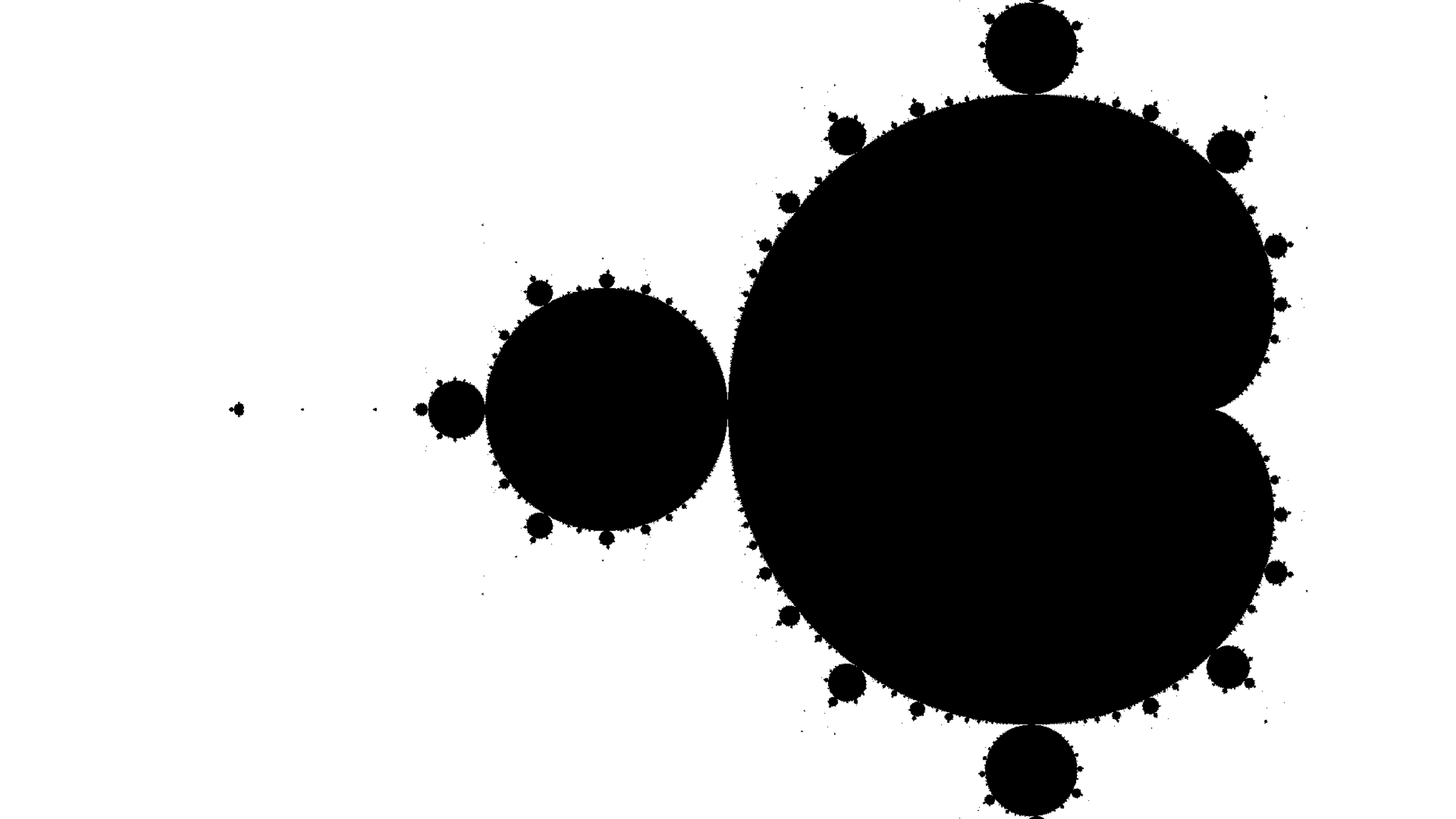}
    \end{center}
    \caption{}
    \label{fig:mandelbrot:shared}
    \end{subfigure}
    \hfill
    \begin{subfigure}[t]{.35\linewidth}
    \begin{center}
    \includegraphics[width=\linewidth,trim={25cm 0 0 0},clip]
    {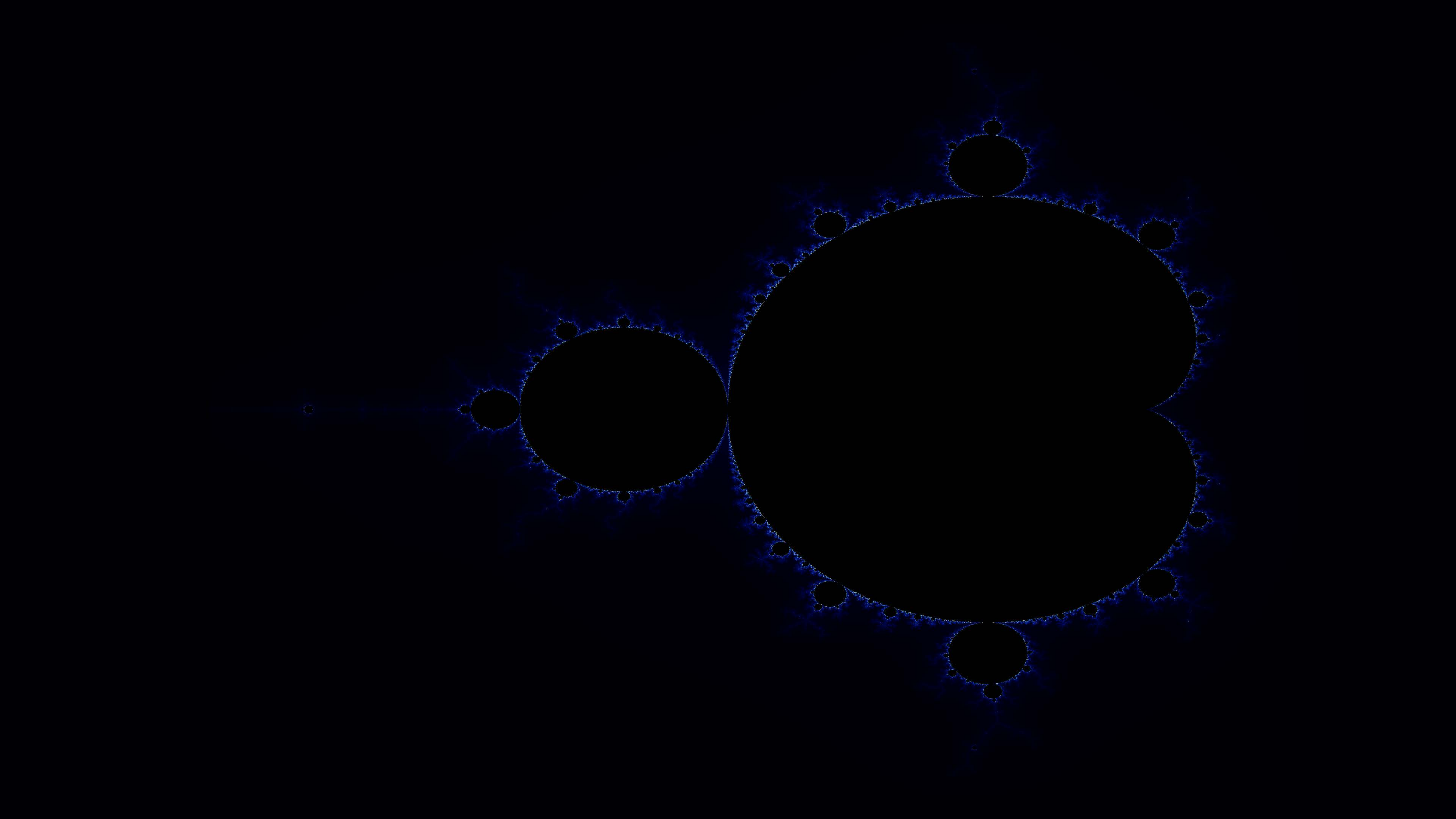}
    \end{center}
    \caption{}
    \label{fig:mandelbrot:dis}
    \end{subfigure}
    
    \caption{Visualization of the Mandelbrot set with $1000$ iterations, scale of $3$, $c_\textbf{real}=-0.75$, and $c_\text{img}=0.0$: (\subref{fig:mandelbrot:shared}) generated by the coroutine code and (\subref{fig:mandelbrot:dis})MPI+OpenMP code using ChatGPT 5. It was quite puzzling that the same model ised different color mapping functions for the Mandelbrot set. For a reference for correctness the first published image of the Mandelbrot set in~\cite{mandelbrot2004fractal} can be used.}
    \label{fig:mandelbrot}
\end{figure*}

The Mandelbrot set was introduced in 1978~\cite{brooks1980dynamics} and first visualized in 1980. The Mandelbrot set of complex numbers $\mathbb{C}$, where the complex number $c \in \mathbb{C}$, for the iterative function reads as
\begin{align}
    z_{n+1} \xrightarrow{} z_n^2 + c, \quad z,c \in \mathbb{C}\text{.}
\end{align}
Figure~\ref{fig:mandelbrot:shared} shows the Mandelbrot set ($1920$ $\times$ $1080$ pixels) plotted for the coroutine code and Figure~\ref{fig:mandelbrot:dis} for the distributed code generated by ChatGPT 5. The default values uses were $1000$ iterations, $c_\textbf{real}=-0.75$, $c_\text{img}=0.0$, and the scale was $3.0$. It was quite puzzling that the same model generated different color mapping functions for the shared memory or distributed memory version of the code. The generated code stored the image in the PBM format and we converted the image to PDF using the \lstinline[language=bash]{convert} tool from ImageMagick\footnote{\url{https://imagemagick.org/}}. For more details, we refer to~\cite[Chapter 4]{diehl2024parallel}. We chose the Mandelbrot set as an artificial workload which is ``embarrassingly parallel''. Another nice aspect is that no solver is required; and the LLM would not need to write a distributed solver, which is quite challenging, or use some library, like eigen or Trilinos.

\section{Prompt design}
\label{sec:prompt:design}

\subsection{Shared memory parallelism}
P1 shows the prompt used for all LLMs to generate the different C\texttt{++} codes. The same prompt was used to generate the asynchronous programming introduced in \cpp~11, the parallel algorithms introduced in \cpp~17, and  coroutines in \cpp~20. To validate the correctness of the images, the requirement to store the image in the PBM format was added. The PBM\footnote{\url{https://en.wikipedia.org/wiki/Netpbm}} format was chosen since it is an easy-to-use text format and no external library is required. For scaling studies the requirement to make the number of threads configurable on the command line was added. Thus, the code would not need to be recompiled for each number of threads.

\begin{displayquote}
\begin{tcolorbox}[colback=gray!15]\textcolor{blue}{P1:}
Write a parallel code for the Mandelbrot set using C\texttt{++} \textbf{\{11 async future,17 parallel algorithms, 20 coroutines\}}  and write the output to the PBM format. Make the number of threads configurable with the command line option \lstinline[language=bash]{-t} and print the runtime excluding IO.
\end{tcolorbox}
\end{displayquote}

P2 shows the prompt provided to all LLMs to generate the code using OpenMP. The prompt is very similar to the one used for the C\texttt{++} code.

\begin{displayquote}
\begin{tcolorbox}[colback=gray!15]
\textcolor{blue}{P2:} Write a parallel code for the Mandelbrot set using C\texttt{++} using OpenMP and write the output to the PBM format. Make the number of threads configurable with the command line option \lstinline[language=bash]{-t} and print the runtime excluding IO.
\end{tcolorbox}
\end{displayquote}

\subsection{Distributed memory parallelism}

P3 shows the prompt provided to all LLMs to generate the \textit{MPI+OpenMP} code.

\begin{displayquote}
\begin{tcolorbox}[colback=gray!15]
\textcolor{blue}{P3:} Write a distributed code using MPI for the Mandelbrot set and using OpenMP for parallelism in C\texttt{++} and write the output to the PBM format. Make the number of partitions configurable with the command line option \lstinline[language=bash]{-p} and print the runtime excluding IO.
\end{tcolorbox}
\end{displayquote}
Additional prompt engineering was required to achieve the correct behavior in the \textit{MPI+OpenMP} example. In particular, the prompt had to explicitly specify that the number of partitions should be configurable. Without this clarification, the generated code redundantly computed the Mandelbrot set multiple times rather than properly distributing the workload.

\section{Results} \label{sec:results}
All codes were tested for compilation using GCC 11.5.0 and build errors were documented. Build errors were fixed by the authors to check for runtime errors. All compiled codes were executed and checked for out-of-bounds errors, segmentation faults, \emph{etc}. Runtime errors were fixed by the authors. The generated PBM file was opened and visually checked for correctness. Figure~\ref{fig:mandelbrot}a shows one correct visualization of the Mandelbrot set. Table~\ref{tab:shared:overview} and Table~\ref{tab:mpi:overview} summarize the build and runtime errors for the shared memory parallelism and distributed memory results, respectively.  (\checkmark) indicates no error and (\tikzxmark) indicates
that errors were found.

\subsection{Shared memory parallelism}

\subsubsection{ChatGPT 4}
For all applications, the following command line options in Listing~\ref{lst:wrong:parameters} were printed by the LLM to change the parameters. However, in all codes only the parameter \lstinline[language=bash]{-t} was parsed and used. All other parameters were ignored and only default values were set. For the scaling study, we had to change the default image size ($1920$ $\times$ $1080$) to make the image larger to scale up to $24$ cores. In addition, the parallel algorithm code set the environment variable \lstinline[language=bash]{OPENMP_NUM_THREADS}; however, GCC uses Intel TBB for parallelism, which ignored the number of threads. The default for Intel TBB is to use all available cores.

\begin{figure*}[b]
    \centering
   \begin{lstlisting}[language=bash,label=lst:wrong:parameters,caption=Command line parameters,commentstyle=\color{asparagus}]
# Usage: [-t threads] [width height [max_iter [x_center y_center [scale]]]]
./exe -t 8 1920 1080 1000 -0.75 0.0 3.0 
\end{lstlisting} 
 
\end{figure*}

\subsubsection{ChatGPT 5}
None of the generated code by ChatGPT 5 had issues, and the issue with the command line arguments by the generated code by ChatGPT 4 was fixed. Furthermore, the correct TBB API call \lstinline[language=c++]{tbb::global_control limit} was used to set the number of threads.

\subsubsection{Claude}
All generated codes compiled. The coroutine code hung and never finished because the \lstinline[language=c++]{while} loop in Listing~\ref{lst:claude:infinite:loop} never terminated, as the \lstinline[language=c++]{done()} function never returned \lstinline[language=c++]{true}. The issue is that true is only returned by \lstinline[language=c++]{std::coroutine_handle<>::done()} when the coroutine is finally suspended. The generated code used \lstinline[language=c++,breaklines=true]{final_suspend=std::suspend_never} and changing that to \lstinline[language=c++,breaklines=true]{final_suspend=std::suspend_always} fixed the issue. All codes produced the correct results. 

\begin{lstlisting}[language=bash,label=lst:claude:infinite:loop,caption=Infinite while loop,commentstyle=\color{asparagus}]
while (!tasks[i].done()) {
    std::this_thread::yield();
}
\end{lstlisting} 

\noindent For the parallel algorithm, the generated code had a comment which mentioned setting the number of threads; see Listing~\ref{lst:claude:threads}. The note is correct, and all threads were used. 

\begin{lstlisting}[language=c++,label=lst:claude:threads,caption=Comment to set the number of threads,commentstyle=\color{asparagus}]
/* Set the number of threads for parallel execution
* Note: C++17 parallel algorithms may not directly use this,
* but some implementations respect thread pool sizes.
*/
\end{lstlisting}

\subsubsection{LLaMA}
The generated code for asynchronous programming produces a black image for a single thread and multiple threads. The parallel algorithm code did not compile with two errors. The first error was 
\lstinline[language=c++]{error: no matching function for call to distance(std::vector<Pixel>::iterator, Pixel*)}. The issue was that the second element of the \lstinline{std::distance} function is supposed to be a \lstinline{std::iterator}. The second issue was \lstinline{policy.execution_policy_tag::concurrency_hint = numThreads;} because \lstinline{execution_policy_tag had not been declared}. After fixing these issues, the code compiled. However, the code ran always using all threads.
For the coroutines, the code did not compile due to a promise mismatch; the code used \lstinline[language=c++]{co_return} and that required \lstinline[language=c++]{Task::promise_type::return_void()}. After fixing that, the code did compile. However, the code stopped with \lstinline[language=bash]{Segmentation fault (core dumped)} while running on a single thread and multiple threads. The issue was that the current work item was released twice. The coroutine released the work item using \lstinline[language=c++]{final_suspend() -> std::suspend_never} and the destructor of the task released it as well. Thus, the second release resulted in a segmentation fault. Once the second release was removed, the code executed. The generated image was all black and the Mandelbrot set was not shown. The issue was that the color was computed for the P5 format; however, it was stored in the P4 format. After fixing that, the correct image was produced. None of the codes set the number of threads correctly and either used all cores or the preset number of cores.

\begin{table}[tb]
    \centering
    \rowcolors{2}{gray!25}{white}
    \begin{tabular}{l|cccc}\toprule
     LLM    &  ChatGPT 5 & ChatGPT 4 & Claude & LLaMA\\\midrule
         & \multicolumn{4}{c}{Coroutines} \\\midrule
    Build   & \checkmark   & \checkmark  & \checkmark & \tikzxmark \\
    Runtime &  \checkmark   & \tikzxmark & \tikzxmark & \tikzxmark  \\
    Correctness & \checkmark & \checkmark & \checkmark & \tikzxmark \\  \midrule
           & \multicolumn{4}{c}{Asynchronous programming} \\\midrule
    Build   & \checkmark   & \checkmark & \checkmark & \checkmark \\
    Runtime &  \checkmark  & \tikzxmark & \checkmark  & \tikzxmark \\
    Correctness & \checkmark & \checkmark & \checkmark & \tikzxmark \\\midrule
    & \multicolumn{4}{c}{Parallel algorithms} \\\midrule
    Build   & \checkmark   & \checkmark & \checkmark & \tikzxmark \\
    Runtime &  \checkmark   &  \tikzxmark & \tikzxmark & \tikzxmark \\
    Correctness & \checkmark & \checkmark & \checkmark & \checkmark \\\midrule
    & \multicolumn{4}{c}{OpenMP} \\\midrule
    Build   & \checkmark   & \checkmark & \checkmark & \checkmark \\
    Runtime & \checkmark    & \tikzxmark & \checkmark & \tikzxmark \\
    Correctness & \checkmark &   \checkmark & \checkmark & \checkmark \\
    \bottomrule
    \end{tabular}
    \caption{Summary of build errors, and compilation errors, and correctness for all generated codes for the shared memory parallelism.}
    \label{tab:shared:overview}
\end{table}

\subsection{Distributed memory}

\subsubsection{ChatGPT 4}
The generated code executed without errors. However, we observed that the implementation used the same color map scheme as the shared memory version, rather than adapting it for the distributed memory context.
\subsubsection{ChatGPT 5}
The generated code executed successfully with no issues encountered during compilation or runtime.

\subsubsection{Claude}
The generated code performed as expected with no errors or issues reported during execution.

\subsubsection{LLaMA}
The generated code compiled, however, the code only executed on a single process. For multiple processes the code crashed with \lstinline[language=bash]{free(): invalid pointer}. After fixing this issue, the code crashed with a segmentation fault using buffer mismatch in \lstinline[language=c++]{MPI_Gather}. 

\subsection{Summary of Results}

\begin{table}[tb]
    \centering
    \rowcolors{2}{gray!25}{white}
    \begin{tabular}{l|cccc}\toprule
     LLM    &  ChatGPT 5 & ChatGPT 4 & Claude & LLaMA\\\midrule
         & \multicolumn{4}{c}{MPI+OpenMP} \\\midrule
    Build   &  \checkmark  & \checkmark   &   \checkmark & \checkmark \\
    Runtime &  \checkmark  & \checkmark &   \checkmark  & \tikzxmark \\
    Correctness  & \checkmark  & \checkmark   & \checkmark &\checkmark \\  
    \bottomrule
    \end{tabular}
    \caption{Summary of build errors, and compilation errors, and correctness for all generated codes for the distributed memory parallelism.}
    \label{tab:mpi:overview}
\end{table}

The success rates for compilation, runtime, and correctness across all evaluated LLM models, for both shared-memory and distributed-memory code, are summarized in Table~\ref{tab:sucess:rate}. For code compilation, all models except LLaMA (40\%) achieved a 100\% success rate. Regarding runtime performance, only ChatGPT-5 achieved a 100\% success rate. Claude reached 40\%, ChatGPT achieved 20\%, and LLaMA successfully executed one out of the generated programs (20\%). In terms of correctness, all models except LLaMA (40\%) achieved a 100\% success rate.

\begin{table}[tb]
    \centering
    \rowcolors{2}{gray!25}{white}
    \begin{tabular}{l|cccc}\toprule
    LLM    & ChatGPT 5 & ChatGPT 4 & Claude & LLaMA \\\midrule
    Build    & 100\% & 100\% & 100\% & 40\%\\
    Runtime & 100\% & 20\% & 40\% & 0\% \\
    Correctness & 100\% & 100\% & 100\% & 40\% \\\bottomrule
    \end{tabular}
    \caption{Success rate for compilation, runtime, and correctness for all models for the shared memory and distributed memory code.}
    \label{tab:sucess:rate}
\end{table}

Table~\ref{tab:llm_errors} summarizes all the errors of the generated codes and the resolution of all errors. For ChatGPT\ 4, the command line parsing error was easy to fix. The library error was more complex since knowledge was needed that Intel TBB is used for parallelism and knowledge about the library on how to set the number of threads. For Claude, the coroutine error was complex since knowledge about \cpp\ coroutines was required. The parallel algorithm bug was easy to spot; however, knowledge that Intel TBB was used was required. The generated code with LLaMA had undefined functions and undeclared variables in asynchronous programming and parallel algorithms, respectively. For the coroutine code, LLaMA had a segmentation fault due to a double release of the current work item, a promise mismatch, and a format mismatch for the PBM format. Most of the bugs were on the simpler side; however, the coroutine-related bugs were on the more complex side, as they required deep knowledge of \cpp\ coroutines.

\begin{table*}[tb]
\centering
\begin{tabular}{lp{2.5cm}p{4.5cm}p{4.5cm}}\toprule
Application & Category & Issue & Resolution \\ \midrule
\multicolumn{4}{c}{ChatGPT\ 4} \\\midrule
All & Command line parsing & Only the parameter \texttt{-t} to set the number of threads was parsed. All other parameters were ignored and default values were used. & Parse the remaining command line options.\\
 \rowcolor{gray!25}Parallel Algo & Library & The code assumed \cpp\ parallel algorithms use OpenMP in GCC for parallelism and \texttt{OPENMP}\texttt{\_NUM}\texttt{\_THREADS} to control the number of threads assuming \cpp parallel algorithms use OpenMP. & \cpp\ parallel algorithms use Intel TBB in GCC for parallelism and \texttt{TBB\_NUM\_THREADS} needs to be set.  \\\midrule
\multicolumn{4}{c}{Claude} \\\midrule
Coroutines & Infinite loop  & \texttt{done()} never returned \texttt{true} because \texttt{final\_suspend} \texttt{=} \texttt{std::suspend\_never} & Changing the suspend action to \texttt{std::suspend\_always} and the loop terminated \\
 \rowcolor{gray!25}Parallel Algo & Parallelism & The code had a comment that the number of thread should be set & Removing the comment and setting the number of threads. \\\midrule
 \multicolumn{4}{c}{LLaMA} \\\midrule
 Async & Undefined function & error: no matching function for call to distance(std::vector$<$Pixel$>$::iterator, Pixel*) & Correct argument to iterator \\
  \rowcolor{gray!25} Parallel Algo & Undeclared variable & execution\_policy\_tag has not been declared & Not using the undeclared variable \\
  Coroutines & Black Mandelbrot set & color was computed for the P5 format, but image was stored in the P4 format & Update the image format to P5 \\
 \rowcolor{gray!25} Coroutines & promise mismatch & code used \texttt{co\_return} &  We had to add to the task the promise type Task::promise\_type::return\_void()  \\
 Coroutines & Segmentation fault & current work item was released twice & Remove the second release of the current work item\\
 \bottomrule
\end{tabular}
\caption{Summary of observed errors for each LLM and their resolutions.}
\label{tab:llm_errors}
\end{table*}

\section{Quality of the generated codes}
\label{sec:code:quality}
Figure~\ref{fig:code:lines:of:code} shows the lines of code (loc) obtained using the \textit{cloc} tool\footnote{\url{https://github.com/AlDanial/cloc}} for the generated code. We observe that in most cases ChatGPT 5 and Claude generated more lines of code than ChatGPT 4 and LLaMA. Only for the generated code using LLaMA did we have to add 39 lines of code to get it working. For all other generated codes no substantial amount of lines of code needed to be added. Table~\ref{tab:code:loc:distributed} shows the lines of code for the distributed codes and similar behavior is observed. To fix the generated code for LLaMA, we had to add 11 lines of code.

\begin{figure*}[tb]
    \centering
\begin{subfigure}[t]{.4\linewidth}
\resizebox{\textwidth}{!}
        {
\begin{tikzpicture}
\begin{axis}[
    xbar stacked,
    xmin=0, xmax=210,
    ytick=data,
    enlarge y limits={abs=1cm},
    symbolic y coords={ChatGPT 5,ChatGPT 4,Claude,LLaMA},
    bar width=10pt,
    xlabel={Lines of Code (LOC)},
    ytick align=outside,
    ytick pos=left,
    major x tick style={transparent},
    legend style={at={(0.7,0.96)},anchor=north west, font=\footnotesize, legend cell align=left},
    xmajorgrids=true
]

% First stack component
\addplot[fill=cadetgrey!60] coordinates {
    (194,ChatGPT 5) 
    (149,ChatGPT 4) 
    (192,Claude) 
    (97,LLaMA)
};

% Second stack component
\addplot[fill=cadetgrey!20] coordinates {
    (0,ChatGPT 5)
    (0,ChatGPT 4)
    (0,Claude)
    (39,LLaMA)
};

\legend{Generated LOC, Fixed LOC}

\end{axis}
\end{tikzpicture}
}
\caption{Coroutines}
\label{fig:code:lines:of:code:coroutine}
\end{subfigure}
\begin{subfigure}[t]{.4\linewidth}
\resizebox{\textwidth}{!}
        {
\begin{tikzpicture}
 \begin{axis}[
    xbar=12pt,
    xmin=0,xmax=160,
    ytick=data,
    enlarge y limits={abs=1cm},
    symbolic y coords={ChatGPT 5,ChatGPT 4,Claude,LLaMA},
    bar width = 10pt,
    xlabel= Lines of code (LOC), 
    ytick align=outside, 
    ytick pos=left,
    major x tick style ={ transparent},
    legend style={at={(0.04,0.96)},anchor=north west, font=\footnotesize, legend cell align=left},
    xmajorgrids=true
        ]    
    \addplot[xbar,fill=cadetgrey!20, area legend] coordinates {
        (109,ChatGPT 5)
        (97,ChatGPT 4)
        (149,Claude)
        (75,LLaMA)
        };
\end{axis}
\end{tikzpicture}
}
\caption{Asynchronous programming}
\label{fig:code:lines:of:code:async}
\end{subfigure}

\begin{subfigure}[t]{.4\linewidth}
\resizebox{\textwidth}{!}
        {
\begin{tikzpicture}
 \begin{axis}[
    xbar=12pt,
    xmin=0,xmax=160,
    ytick=data,
    enlarge y limits={abs=1cm},
    symbolic y coords={ChatGPT 5,ChatGPT 4,Claude,LLaMA},
    bar width = 10pt,
    xlabel= Lines of code (LOC), 
    ytick align=outside, 
    ytick pos=left,
    major x tick style ={ transparent},
    legend style={at={(0.04,0.96)},anchor=north west, font=\footnotesize, legend cell align=left},
    xmajorgrids=true
        ]    
    \addplot[xbar,fill=cadetgrey!20, area legend] coordinates {
        (119,ChatGPT 5)
        (75,ChatGPT 4)
        (151,Claude)
        (73,LLaMA)
        };
\end{axis}
\end{tikzpicture}
}
\caption{Parallel Algorithms}
\label{fig:code:lines:of:code:par}
\end{subfigure}
\begin{subfigure}[t]{.4\linewidth}
\resizebox{\textwidth}{!}
        {
\begin{tikzpicture}
 \begin{axis}[
    ybar stacked,
    xbar=12pt,
    xmin=0,xmax=140,
    ytick=data,
    enlarge y limits={abs=1cm},
    symbolic y coords={ChatGPT 5,ChatGPT 4,Claude,LLaMA},
    bar width = 10pt,
    xlabel= Lines of code (LOC), 
    ytick align=outside, 
    ytick pos=left,
    major x tick style ={ transparent},
    legend style={at={(0.04,0.96)},anchor=north west, font=\footnotesize, legend cell align=left},
    xmajorgrids=true
        ]    
    \addplot[xbar,fill=cadetgrey!20, area legend] coordinates {
        (104,ChatGPT 5)
        (83,ChatGPT 4)
        (137,Claude)
        (68,LLaMA)
        };
\end{axis}

\end{tikzpicture}
}
\caption{OpenMP}
\label{fig:code:lines:of:code:openmp}
\end{subfigure}
    \caption{Lines of code (LOC) for (\protect\subref{fig:code:lines:of:code:coroutine}) coroutines, (\protect\subref{fig:code:lines:of:code:async}) asynchronous programming, (\protect\subref{fig:code:lines:of:code:par}) parallel algorithms, and (\protect\subref{fig:code:lines:of:code:openmp}) OpenMP; generated by the AI model including comments for all four shared memory examples. We used the tool \textit{cloc} to obtain the lines of code. Note the the plot of lines of code for coroutines (\protect\subref{fig:code:lines:of:code:coroutine}) is different since this was the only generated code where we had to add 37 lines of code to fix the LLaMA generated code. For all other codes the fixed were smaller and mostly existing line of codes had to be changed.}
    \label{fig:code:lines:of:code}
\end{figure*}

\begin{table}[tb]
    \centering
    \begin{tabular}{l|cccc}\toprule
    LLM     & ChatGPT 5 & ChatGPT 4 & Claude & LLaMA  \\\midrule
    LOC     & 143 & 91 & 116 & 71/\textcolor{azure}{82} \\\bottomrule
    \end{tabular}
    \caption{Lines of code (LOC) generated by the AI model including comments for all distributed memory examples. The \textcolor{azure}{blue} numbers are lines of code after the correction. We used the tool \textit{cloc} to obtain the lines of code. }
    \label{tab:code:loc:distributed}
\end{table}

To quantify how difficult it was to develop the code, we used the \textbf{Co}nstructive \textbf{Co}st \textbf{Mo}del (COCOMO)~\cite{5010193,1237981}. The COCOMO model aims for serial code and does not take parallel or distributed programming into account. An attempt to investigate parallel programming was made with the COCOMO \textit{II} model~\cite{miller2018applicability}. Starting in the 90s, the HPC community discussed having a cost model for HPC codes; however, no model has been proposed as of today. Therefore, we use the COCOMO model. We used the open-source tool \textit{scc}\footnote{\url{https://github.com/boyter/scc}} to get the COCOMO metrics.

\begin{figure}[tb]
    \centering
    \begin{tikzpicture}[scale=0.95]
    \draw[help lines, color=gray!30, dashed] (-0.1,-0.1) grid (2.9,2.9);
    \draw[->,thick,cadetgrey] (0,0)--(3,0) node[right]{Difficult};
    \draw[->,thick,cadetgrey] (0,0)--(0,3.1) node[above,cadetgrey]{Good};
    \node[left,cadetgrey] at (0,0) {Easy};
    \node[below,cadetgrey] at (0,0) {Poor};
    % Coroutines
    \node[below,black] at (3,3) {\textbf x};
    \node[above,azure] at (2.701657459,2) {\textbf x};
    \node[below,asparagus] at (2.983425414,2) {\textbf x};
    \node[above,cadetgrey] at (2.602209945,0) {\textbf x};
    % OpenMP
    \node[below,black] at (2.337016575,3) {$\bigstar$};
    \node[above,azure] at (2.138121547,2) {$\bigstar$};
    \node[below,asparagus] at (2.61878453,3) {$\bigstar$};
    \node[above,cadetgrey] at (1.972375691,2) {$\bigstar$};
    % async
    \node[above,black] at (2.386740331,3) {$\spadesuit$};
    \node[below,azure] at (2.270718232,2) {$\spadesuit$};
    \node[above,asparagus] at (2.701657459,3) {$\spadesuit$};
    \node[above,cadetgrey] at (2.055248619,1) {$\spadesuit$};
    % parallel
    \node[left,black] at (2.46961326,3) {$\clubsuit$};
    \node[below,azure] at (2.055248619,2) {$\clubsuit$};
    \node[above,asparagus] at (2.718232044,3) {$\clubsuit$};
    \node[above,cadetgrey] at (2.038674033,1) {$\clubsuit$};
    % Legend
    \draw (4.4,3.3) -- (8.5,3.3) -- (8.5,1.2) -- (4.4,1.2) -- cycle;
    \node[right,cadetgrey] at (4.5,3) {$\textbf x$ Coroutines};
    \node[right,cadetgrey] at (4.5,2.5) {$\spadesuit$ Async programming};
    \node[right,cadetgrey] at (4.5,2) {$\clubsuit$ Parallel Algorithms};
     \node[right,cadetgrey] at (4.5,1.5) {$\bigstar$ OpenMP};
    \end{tikzpicture}
    \caption{Quality from poor to good and the effort from easy to difficult obtained by the COCOMO model using the estimated months. In black ChatGPT 5, in \textcolor{azure}{blue} ChatGPT 4, in \textcolor{asparagus}{green} Claude, and \textcolor{cadetgrey}{grey} Lamma.}
    \label{fig:effort:shared}
\end{figure}
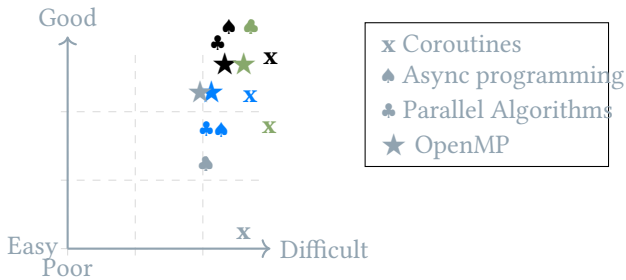

\noindent Figure~\ref{fig:effort:shared} shows the effort and quality of the generated codes for the shared memory parallelism. We observe that LLaMA has two outliers for the coroutines and parallel algorithms. ChatGPT 5 performed the best and all results are in the right upper corner. Claude performed the second best and ChatGPT 4 third. From the estimated efforts, all generated codes behaved similarly. 
% Section~\ref{sec:scaling} evaluates the scaling behavior. Finally, Section~\ref{sec:conclusion} concludes the paper. 

\begin{figure}[tb]
    \centering
    \begin{tikzpicture}[scale=0.95]
    \draw[help lines, color=gray!30, dashed] (-0.1,-0.1) grid (2.9,2.9);
    \draw[->,thick,cadetgrey] (0,0)--(3,0) node[right]{Difficult};
    \draw[->,thick,cadetgrey] (0,0)--(0,3.1) node[above,cadetgrey]{Good};
    \node[left,cadetgrey] at (0,0) {Easy};
    \node[below,cadetgrey] at (0,0) {Poor};
    % ChatGPT4
    \node[below,black] at (2.5125,3) {\textbf x};
    % LLaMA
    \node[below,black] at (2.41875,2) {$\bigstar$};
    % ChatGPt 5
    \node[above,black] at (3,3) {$\spadesuit$};
    % Claude
    \node[left,black] at (2.775,3) {$\clubsuit$};
    % Legend
    \draw (4.4,3.3) -- (8,3.3) -- (8,1.2) -- (4.4,1.2) -- cycle;
    \node[right,cadetgrey] at (4.5,3) {$\textbf x$ ChatGPT 4};
    \node[right,cadetgrey] at (4.5,2.5) {$\spadesuit$ ChatGPT 5};
    \node[right,cadetgrey] at (4.5,2) {$\clubsuit$ Claude};
     \node[right,cadetgrey] at (4.5,1.5) {$\bigstar$ LLaMA};
    \end{tikzpicture}
    \caption{Quality from poor to good and the effort from easy to difficult obtained by the COCOMO model using the estimated months. 
    %In black ChatGPT 5, in \textcolor{azure}{blue} ChatGPT 4, in \textcolor{asparagus}{green} Claude, and \textcolor{cadetgrey}{grey} Lamma.
    }
    \label{fig:effort:distributed}
\end{figure}
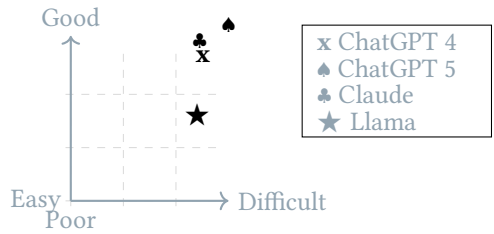

\noindent Figure~\ref{fig:effort:distributed} shows the effort and quality of the generated codes for the distritbuted memory parallelism. The ChatGPT 5 code had more effort than the other codes. The Claude and ChatGPT 4 generated codes have similar efforts. The LLaMA code had less effort and was the only code with runtime issues.

\section{Scaling results}
\label{sec:scaling}
The scaling results were obtained on the rostam cluster at Louisiana State University (LSU). The medusa partition has 16 nodes with one Intel\textsuperscript{\sffamily\textregistered}\  Xeon\textsuperscript{\sffamily\textregistered}\ Gold 6148 CPU @ 2.40GHz (20 cores) and 92 GB memory. GCC 11.5.0 and OpenMPI 5.0.7 were used to compile the generated code. For the parallel algorithms Intel\textsuperscript{\sffamily\textregistered}\ TBB 2022.2.

\subsection{Shared memory}
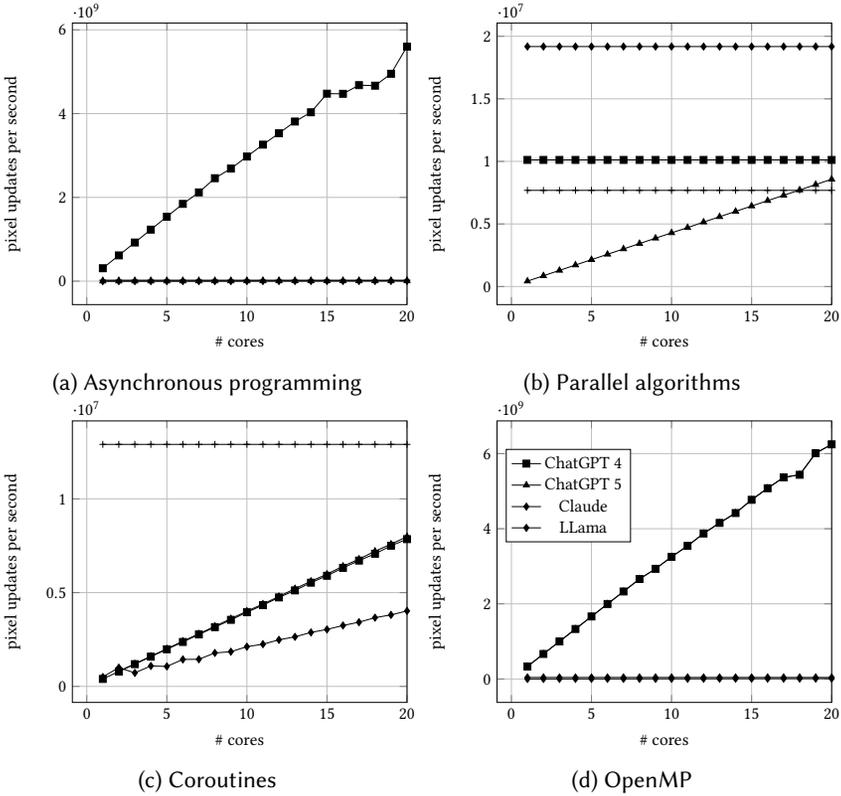
\begin{figure*}[tb]
\centering
\begin{subfigure}[t]{0.4\textwidth}
\begin{tikzpicture}[scale=0.65, transform shape]
\begin{axis}[xlabel=\# cores,ylabel=pixel updates per second,grid,xmax=20]
\addplot[black,mark=square*] table [x expr=\thisrowno{0},y expr={2073600000/(\thisrowno{1}*0.001)}, col sep=comma] {data_async_gpt4.csv};
\addplot[black,mark=triangle*] table [x expr=\thisrowno{0},y expr={2073600000/(\thisrowno{1}*0.001)}, col sep=comma] {data_async_gpt5.csv};
\addplot[black,mark=diamond*] table [x expr=\thisrowno{0},y expr={2073600000/(\thisrowno{1}*0.001)}, col sep=comma] {data_async_claude.csv};
\addplot[black,mark=o] table [x expr=\thisrowno{0},y expr={2073600000/(\thisrowno{1})}, col sep=comma] {data_async_LLaMA.csv};
\addplot[azure,mark=heart]  table [x index = {0}, y expr={8294400000/\thisrowno{1}}, col sep=comma] {data_benchmark_future.csv};
\end{axis}
\end{tikzpicture}
\caption{Asynchronous programming}
\label{fig:scaling:shared:async}
\end{subfigure}
\begin{subfigure}[t]{0.4\textwidth}
\begin{tikzpicture}[scale=0.65, transform shape]
\begin{axis}[xlabel=\# cores,ylabel=pixel updates per second,grid,xmax=20]
\addplot[black,mark=square*] table [x expr=\thisrowno{0},y expr={2073600000/\thisrowno{1}}, col sep=comma] {data_paralgo_gpt4.csv};
\addplot[black,mark=triangle*] table [x expr=\thisrowno{0},y expr={2073600000/(\thisrowno{1}*0.001)}, col sep=comma] {data_paralgo_gpt5.csv};
\addplot[black,mark=diamond*] table [x expr=\thisrowno{0},y expr={2073600000/(\thisrowno{1}*0.001)}, col sep=comma] {data_paralgo_claude.csv};
\addplot[black,mark=o] table [x expr=\thisrowno{0},y expr={2073600000/(\thisrowno{1})}, col sep=comma] {data_paralgo_LLaMA.csv};
\addplot[azure,mark=heart]  table [x index = {0}, y expr={829440000/\thisrowno{1}}, col sep=comma] {data_benchmark_algorithms.csv};
\end{axis}
\end{tikzpicture}
\caption{Parallel algorithms}
\label{fig:scaling:shared:paralgo}
\end{subfigure}

\begin{subfigure}[t]{0.4\textwidth}
\begin{tikzpicture}[scale=0.65, transform shape]
\begin{axis}[xlabel=\# cores,ylabel=pixel updates per second,grid,xmax=20]
\addplot[black,mark=square*] table [x expr=\thisrowno{0},y expr={2073600000/\thisrowno{1}}, col sep=comma] {data_coroutines_gpt4.csv};
\addplot[black,mark=triangle*] table [x expr=\thisrowno{0},y expr={2073600000/(\thisrowno{1}*0.001)}, col sep=comma] {data_coroutines_gpt5.csv};
\addplot[black,mark=diamond*] table [x expr=\thisrowno{0},y expr={2073600000/(\thisrowno{1}*0.001)}, col sep=comma] {data_coroutines_claude.csv};
\addplot[black,mark=o] table [x expr=\thisrowno{0},y expr={2073600000/(\thisrowno{1}*0.001)}, col sep=comma] {data_coroutines_LLaMA.csv};
\addplot[azure,mark=heart]  table [x index = {0}, y expr={829440000/\thisrowno{1}}, col sep=comma] {data_benchmark_coroutine.csv};
\end{axis}
\end{tikzpicture}
\caption{Coroutines}
\label{fig:scaling:shared:coroutines}
\end{subfigure}
\begin{subfigure}[t]{0.4\textwidth}
\begin{tikzpicture}[scale=0.65, transform shape]
\begin{axis}[xlabel=\# cores,ylabel=pixel updates per second,grid,xmax=20,legend style={at={(0.4,0.9)},anchor=north east}]
\addplot[black,mark=square*] table [x expr=\thisrowno{0},y expr={2073600000/(\thisrowno{1}*0.001)}, col sep=comma] {data_openmp_gpt4.csv};
\addplot[black,mark=triangle*] table [x expr=\thisrowno{0},y expr={2073600000/(\thisrowno{1}*0.001)}, col sep=comma] {data_openmp_gpt4.csv};
\addplot[black,mark=diamond*] table [x expr=\thisrowno{0},y expr={2073600000/(\thisrowno{1}*0.001)}, col sep=comma] {data_openmp_claude.csv};
\addplot[black,mark=o] table [x expr=\thisrowno{0},y expr={2073600000/(\thisrowno{1})}, col sep=comma] {data_openmp_LLaMA.csv};
\addplot[azure,mark=heart]  table [x index = {0}, y expr={829440000/\thisrowno{1}}, col sep=comma] {data_benchmark_openmp.csv};
\addlegendentry{ChatGPT 4}
\addlegendentry{ChatGPT 5}
\addlegendentry{Claude}
\addlegendentry{LLaMA}
\addlegendentry{Human~\cite{diehl2024parallel}}
\end{axis}
\end{tikzpicture}
\caption{OpenMP}
\label{fig:scaling:shared:openmp}
\end{subfigure}
    \caption{Single node scaling: (\subref{fig:scaling:shared:async}) asynchronous programming, (\subref{fig:scaling:shared:paralgo}) parallel programming, (\subref{fig:scaling:shared:coroutines}) coroutines, and (\subref{fig:scaling:shared:openmp}) OpenMP. The legend is the same for all plots but only shown in the last figure. Note that the straight lines in some cases indicate scaling, however, the processed pixels per second are one magnitude lower. In other cases the number of threads was set to all and the generated code ignored the number of threads. As a reference we plotted in \textcolor{azure}{blue} the results of human generated codes. These codes were used in the textbook ``
Parallel C\texttt{++}: Efficient and Scalable High-Performance Parallel Programming Using HPX
''~\cite{diehl2024parallel}. Note we used a heart to indicate the human code developer.}
    \label{fig:scaling:shared}
\end{figure*}

Figure~\ref{fig:scaling:shared} shows the scaling observed on a single node. The scaling for the asynchronous programming is shown in Figure~\ref{fig:scaling:shared:async}. ChatGPT 4 scaled up to 15 cores linearly; after that, scaling was still observed. ChatGPT 5 scaled up to 20 cores; however, the pixels processed per second are one order of magnitude lower and therefore a straight line. Claude scaled up to 20 cores; however, the pixels processed per second are one order of magnitude lower and therefore appear as a straight line. LLaMa generated code which ignored the number of threads and always ran on all cores. Figure~\ref{fig:scaling:shared:paralgo} shows the scaling for parallel algorithms. The code generated by ChatGPT 4 always used all 20 cores and therefore appear as a straight line. ChatGPT 5 fixed the issue of the ChatGPT 4 generated code and the number of cores was set correctly. Therefore, the code scaled with the number of cores. Claude ignored the number of threads as well and ran on all available cores. Therefore, we see a straight line. LLaMA generated code that ignored the number of threads and always ran on all cores. Figure~\ref{fig:scaling:shared:coroutines} shows the scaling for coroutines. The code generated by ChatGPT 4 and ChatGPT 5 showed similar scaling. The code generated by Claude scaled too; however, it was a little bit slower as the ChatGPT codes. Figure~\ref{fig:scaling:shared:openmp} shows the scaling for OpenMP. For ChatGPT 4 and ChatGPT 5 the generated code scaled. For Claude, the generated code scaled; however, the processed pixels per second were one order of magnitude lower and therefore a straight line is shown. LLaMA generated code, which ignored the number of threads, always ran on all cores. However, the code was one order of magnitude slower.

One important aspect is to compare the LLM generated code with the human generated codes. The code presented in the textbook ``
Parallel C\texttt{++}: Efficient and Scalable High-Performance Parallel Programming Using HPX
''~\cite{diehl2024parallel} for the same examples is used as a comparison. The codes are available on GitHub~\footnote{\url{https://github.com/ModernCPPBook/Examples}}. We decided to use codes from a textbook to have optimized codes, but not highly optimized ones. For asynchronous programming, ChatGPT 4 added some optimizations and outperformed the textbook version of the code. For the parallel algorithms, the human generated code scaled and outperformed the only scaling code from ChatGPT 5. For coroutines, the human generated code outperformed ChatGPT 4/5 and Claude. However, the LLaMA generated code did not consider the number of cores but was faster when using all cores. For OpenMP, ChatGPT 5 generated a code outperforming all other LLM generated codes and the human code.

\subsection{Distributed memory}
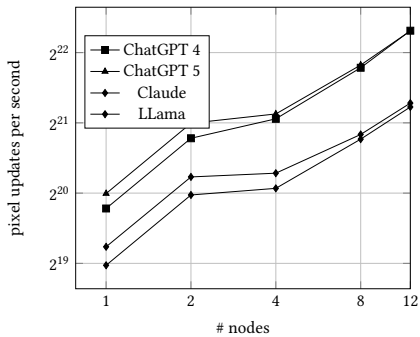
\begin{figure}[tb]
    \centering
    \begin{tikzpicture}[scale=0.9, transform shape]
\begin{axis}[xlabel=\# nodes,ylabel=pixel updates per second,grid,xmax=12,legend style={at={(0.4,0.95)},anchor=north east},xtick={1,2,4,8,12},xmode=log,log basis x=2,ymode=log,log basis y=2,xticklabels={1,2,4,8,12}]
\addplot[black,mark=square*] table [x expr=\thisrowno{0},y expr={960000000/(\thisrowno{1})}, col sep=comma] {data_mpi_gpt4.csv};
\addplot[black,mark=triangle*] table [x expr=\thisrowno{0},y expr={960000000/(\thisrowno{1})}, col sep=comma] {data_mpi_gpt5.csv};
\addplot[black,mark=diamond*] table [x expr=\thisrowno{0},y expr={960000000/(\thisrowno{1})}, col sep=comma] {data_mpi_claude.csv};
\addplot[black,mark=*] table [x expr=\thisrowno{0},y expr={960000000/(\thisrowno{1})}, col sep=comma] {data_mpi_LLaMA.csv};
\addplot[azure,mark=heart]  table [x index = {0}, y expr={829440000/\thisrowno{1}}, col sep=comma] {data_mpi.csv};
\addlegendentry{ChatGPT 4}
\addlegendentry{ChatGPT 5}
\addlegendentry{Claude}
\addlegendentry{LLaMA}
\end{axis}
\end{tikzpicture}
    \caption{Distributed scaling from a single node up to 12 nodes. As a reference we plotted in \textcolor{azure}{blue} the results of human generated codes. These codes were used in the textbook ``
Parallel C\texttt{++}: Efficient and Scalable High-Performance Parallel Programming Using HPX
''~\cite{diehl2024parallel}. Note we used a heart to indicate the human code developer.}
    \label{fig:scaling:distributed}
\end{figure}

For the distributed runs, the pixels were set to $(12000 \times 80000)$, and $1000$ iterations were used. Figure~\ref{fig:scaling:distributed} shows the scaling results from a single node up to 12 nodes (1,2,4,8, and 12). We used one MPI rank per node and 20 cores per node. The image was partitioned such that each MPI rank computed one partition. For ChatGPT 4, the code scaled; however, the speedup from two to four nodes was lower. For ChatGPT 5, the code scaled; however, the speedup from two to four nodes was lower. The performance from 8 to 12 nodes was similar for both OpenAI models. For Claude, the code scaled as well, but was slower compared to the OpenAI models. For LLaMA, the code placed the time measurement code for the end time incorrectly and excluded the MPI gathering. We had to fix that to be aligned with the other codes. For LLaMA, the code scaled and had comparable performance to Claude. Interestingly, all codes had the scaling issues when going from two to four nodes. Generally, it was quite impressive that all the generated code scaled after fixing some runtime issues.

One important aspect is to compare the LLM generated with human generated codes. The code presented in the textbook ``
Parallel C\texttt{++}: Efficient and Scalable High-Performance Parallel Programming Using HPX
''~\cite{diehl2024parallel} for the same examples is used as a comparison. The codes are available on GitHub~\footnote{\url{https://github.com/ModernCPPBook/Examples}}. We decided to use codes from a textbook to have optimized codes, but not highly optimized ones. For the distributed codes, the human generated code outperformed all LLM generated codes. A sophisticated analysis of the different communication patterns is needed to provide more explanation.  

\section{Conclusion and Outlook}
\label{sec:conclusion}
In conclusion, this study provides a comprehensive evaluation of the capabilities and limitations of state-of-the-art LLMs in generating high-performance parallel and distributed C\texttt{++} code for computing the Mandelbrot set. Our analysis demonstrates that while LLMs such as ChatGPT 4 and 5 exhibit strong syntactic accuracy and the ability to produce functionally correct and scalable code across multiple parallel paradigms, their performance consistency and efficiency still lag behind expert-crafted implementations. Claude and LLaMA models, though able to produce runnable code, often failed to manage thread control and optimization parameters effectively, leading to suboptimal scaling and performance degradation.

Understanding why some models perform well while others do not, remains challenging. LLMs largely function as black boxes: their training data, model architectures, and optimization strategies are not fully disclosed. In the context of code generation, it is reasonable to assume that training data includes large amounts of publicly available source code, such as repositories hosted on GitHub. However, it is far less clear whether—and to what extent—specialized data related to thread management and distributed synchronization was included in the training process. Model scale is another important but opaque factor. Companies generally do not publish exact parameter counts, making direct comparisons difficult. Based on external estimates, ChatGPT-5 is believed to have on the order of 600 billion parameters, while smaller models are estimated at around 31 billion parameters. Claude is estimated to have approximately 175 billion parameters, and LLaMA around 400 billion parameters. These differences in scale, combined with unknown training data and objectives, likely contribute significantly to the observed variation in model performance.

Despite these limitations in understanding the internal workings of LLMs, our findings demonstrate both the promise and current boundaries of AI-assisted code synthesis in HPC contexts. The variability in model performance—particularly in thread coordination, memory management, and scaling behavior—suggests that future improvements will require not only larger models or more data, but potentially more targeted training on HPC-specific patterns and paradigms. Ultimately, this work establishes a foundational benchmark for assessing LLM-driven HPC code generation, paving the way for future advancements in autonomous scientific computing and intelligent parallel programming assistance.

% The findings underscore both the potential and current boundaries of AI-assisted code synthesis in HPC contexts, highlighting the need for improved understanding of memory models, thread coordination, and compiler interactions within LLM architectures. . 

\subsection{Outlook}
Future work will extend this evaluation to communication-intensive benchmarks, such as stencil computations with halo exchange, parallel FFTs, and iterative solvers relying on collective MPI operations, in order to further stress-test LLM-generated code under more demanding and realistic HPC conditions. Another important direction is the inclusion of more complex scientific applications, for example finite element or finite difference methods for computational mechanics, as well as other representative workloads.

For comparisons with human-generated code, more highly optimized implementations could be employed to better contextualize and validate performance outliers observed in some LLM-generated results. In addition, a more sophisticated analysis—leveraging profiling and debugging tools—would be valuable for gaining deeper insight into why certain LLM-generated codes perform significantly better than others. Such an investigation, however, is beyond the scope of this paper.

Currently, this study evaluates only ChatGPT, Claude, and LLaMA-based models. To provide a more comprehensive and representative assessment of modern large language models for code generation, future work should include additional state-of-the-art systems such as Google’s Gemini family, DeepSeek-Coder, Grok, and other specialized code-oriented models.

Furthermore, it would be valuable to compare LLM-based approaches with established non-LLM developer tools, such as Intel® Parallel Advisor and similar performance analysis and optimization frameworks. Including such tools would provide a broader perspective on how generative AI systems compare to traditional static analysis, profiling, and parallelization technologies in practical software engineering workflows.

\section*{Supplementary materials}
The generated code is available on GitHub\footnote{\url{https://github.com/diehlpkpapers/distributed-cpp-ai}} and Zenodo~\cite{diehl_2026_18895900}, respectively. The human generated codes from the textbook~\cite{diehl2024parallel} are available on GitHub\footnote{\url{https://github.com/ModernCPPBook/Examples}}.

\section*{Acknowledgment}
\footnotesize
This work was supported by the U.S. Department of Energy through the Los Alamos National Laboratory. Los Alamos National Laboratory is operated by Triad National Security, LLC, for the National Nuclear Security Administration of U.S. Department of Energy (Contract No. 89233218CNA000001). Approved by LANL as LA-UR-25-30538. 

\bibliographystyle{IEEEtran}
\bibliography{References}

\end{document}